\documentclass[%
 aip,
 amsmath,amssymb,
 reprint,%
]{revtex4-1}

\usepackage{graphicx}
\usepackage{dcolumn}
\usepackage{bm}

\usepackage[utf8]{inputenc}
\usepackage[T1]{fontenc}
\usepackage{mathptmx}
\usepackage{etoolbox}
\usepackage{xcolor}

\makeatletter
\def\@email#1#2{%
 \endgroup
 \patchcmd{\titleblock@produce}
  {\frontmatter@RRAPformat}
  {\frontmatter@RRAPformat{\produce@RRAP{*#1\href{mailto:#2}{#2}}}\frontmatter@RRAPformat}
  {}{}
}%
\makeatother
\begin{document}

\preprint{AIP/123-QED}
\title[Measurement of Spin-Polarized Photoemission from Wurtzite and Zinc-Blende Gallium Nitride Photocathodes]{Measurement of Spin-Polarized Photoemission from Wurtzite and Zinc-Blende Gallium Nitride Photocathodes}
\author{S.J. Levenson}
 \email{sjl354@cornell.edu.}
 \author{M.B. Andorf}
  \author{B.D. Dickensheets}  
   
      
        \author{I.V. Bazarov}
         
\affiliation{ 
Cornell Laboratory for Accelerator-Based Sciences and Education (CLASSE), Cornell University, Ithaca, NY 14850, USA}%

\author{A. Galdi}
\affiliation{Department of Industrial Engineering, University of Salerno, Fisciano (SA) 84084, Italy}

 \author{J. Encomendero}
\author{V.V. Protasenko}
\affiliation{ 
School of Electrical and Computer Engineering, Cornell University, Ithaca, New York 14853, USA}%

\author{D. Jena}
\author{H.G. Xing}
\affiliation{ 
School of Electrical and Computer Engineering, Cornell University, Ithaca, New York 14853, USA}%
\affiliation{ 
Department of Material Science and Engineering, Cornell University, Ithaca, NY 14583, USA}%
\affiliation{ 
Kavli Institute at Cornell for Nanoscale Science, Cornell University, Ithaca, NY 14583, USA}%
\author{J.M. Maxson}
\affiliation{ 
Cornell Laboratory for Accelerator-Based Sciences and Education (CLASSE), Cornell University, Ithaca, NY 14850, USA}%

\date{\today}

\begin{abstract}
Spin-polarized photoemission from wurtzite and zinc-blende gallium nitride (GaN) photocathodes has been observed and measured for the first time. The p-doped GaN photocathodes were epitaxially grown and activated to negative electron affinity (NEA) with a cesium monolayer deposited on their surfaces. A field-retarding Mott polarimeter was used to measure the spin-polarization of electrons photoemitted from the top of the valence band. A spectral scan with a tunable optical parametric amplifier (OPA) constructed to provide low-bandwidth light revealed peak spin polarizations of 17\% and 29\% in the wurtzite and zinc-blende photocathodes, respectively. Zinc-blende GaN results are analyzed with a spin-polarization model accounting for experimental parameters used in the measurements, while possible mechanisms influencing the obtained spin polarization values of wurtzite GaN are discussed.

\end{abstract}

\maketitle

Spin-polarized electron beam generation is a critical component for many nuclear and particle physics experiments by enabling the study of spin-dependent properties such as parity-violation and nucleon spin structure \cite{nuclearpaper, ncpaper1, ncpaper2, ncpaper3}. Spin-polarized electrons are also valuable in electron microscopy for investigating magnetic structures and material spin states \cite{microscopypaper, microscopypaper2, microscopypaper3}. For these applications, photocathodes based on Gallium Arsenide (GaAs) are currently the only viable spin-polarized source and are employed at facilities such as the Mainz Microtron (MAMI) \cite{mami} and the Continuous Electron Beam Accelerator Facility (CEBAF) \cite{cebaf}. Future facilities like the International Linear Collider (ILC) \cite{nuclearpaper} and Electron-Ion Collider (EIC) \cite{eicpaper}, which call for higher average beam currents than facilities operating today, can benefit from a more robust spin-polarized electron source. 

 Spin-polarized electrons from GaAs\cite{piercepaper} are produced with circularly-polarized light at a photon energy tuned to the material bandgap to photoexcite electrons occupying the top of the valence band into the conduction band. Ordinarily, these electrons could not contribute to the photoemission process because their energy is not sufficient to overcome the workfunction. Fortunately, GaAs can be brought to a condition called Negative Electron Affinity (NEA) where the conduction band minimum is at a higher energy than the vacuum on the surface, thus enabling bandgap energy photons to produce photoemission. Bulk GaAs is limited to a theoretical maximum spin-polarization of 50\% because of its light-hole (LH) and heavy-hole (HH) degenerate band states \cite{pierce2}, while measured polarization is around 35\% due to various depolarization mechanisms\cite{BAP, DYP, EY, opaper}. In strained superlattice GaAs, the degeneracy can be removed and $>$90\% spin polarizations are achievable \cite{superlatticepaper}. 

To activate GaAs to NEA, about one cesium monolayer and an oxidant are deposited on the surface \cite{oldneapaper}. This layer is highly reactive to gas molecules and, therefore, GaAs operation requires an extreme-high vacuum (XHV) environment, which restricts its use to DC electron guns only. Despite XHV operation, chemical poisoning \cite{chemicalpoisoningpaper1, chemicalpoisoningpaper2} along with ion back-bombardment \cite{ibbpaper} and thermal desorption \cite{thermalpaper} severely limit the operational lifetime of NEA GaAs photocathodes. While much work has been done to improve the lifetime \cite{j1paper, j2paper, j3paper, j4paper, j5paper}, there is motivation to investigate other materials as potential robust spin-polarized electron beam sources\cite{lucarev}. A recent result indicates an alkali-antimonide photocathode (K$_2$NaSb) can produce spin-polarized electrons \cite{aaspin}. In this work we present the first ever spin-polarized electron beam demonstration from  Gallium Nitride (GaN) based photocathodes grown in both the hexagonal wurzite (W-GaN) and cubic zinc-blende (ZB-GaN) crystal orientations \cite{ipac23paper}.

GaN is a direct bandgap material widely studied for its use in light emitting diodes (LEDs) and other opto- and micro-electronics \cite{gan0,gan1, ganext, gan2, gan3, gan4}. W-GaN\cite{singlecrystalgan} and ZB-GaN\cite{cubicgan} have respective bandgap energies ($E_g$) of 3.4 and 3.2 eV \cite{semicondbook}. Like GaAs, a GaN photocathode requires activation to NEA for efficient operation. Despite its common use in the electronics industry, GaN has been far less investigated as a photocathode material because growth techniques for high-quality samples were developed much more recently than those of GaAs \cite{ganreview}. Nonetheless, GaN boasts many interesting photocathode properties \cite{ivanpaper, ipac22paper}. P-doped W-GaN has achieved $>$50\% quantum efficiency (QE)\cite{highqepaper} and N-polar GaN was used to achieve NEA without Cs \cite{Npolar1}. The thermal conductivity of GaN is more than a factor of two larger than that of GaAs \cite{thermalcondpapergaas, thermalcondpapergan} which may mitigate thermal desorption effects at high operational beam currents. NEA W-GaN has also demonstrated a 20 times longer dark lifetime at its emission threshold energy of 365 nm compared to GaAs at 780 nm \cite{thresholdpaper}, indicating GaN to be a more robust photo-emitter. GaN has been extensively studied for its applicability in spintronics and spin-polarized devices \cite{spinlaser, strainedpol1, inganpol,wganpoluse4, wganpoluse1}. Several works have reported asymmetries between left and right circularly-polarized light emission in W-GaN resulting from spin-polarized excitation \cite{wganpoluse1,wganpoluse2, wganpoluse3,  bhattacharya}. 

\begin{figure}[h]
   \centering
   \includegraphics*[width=1\columnwidth]{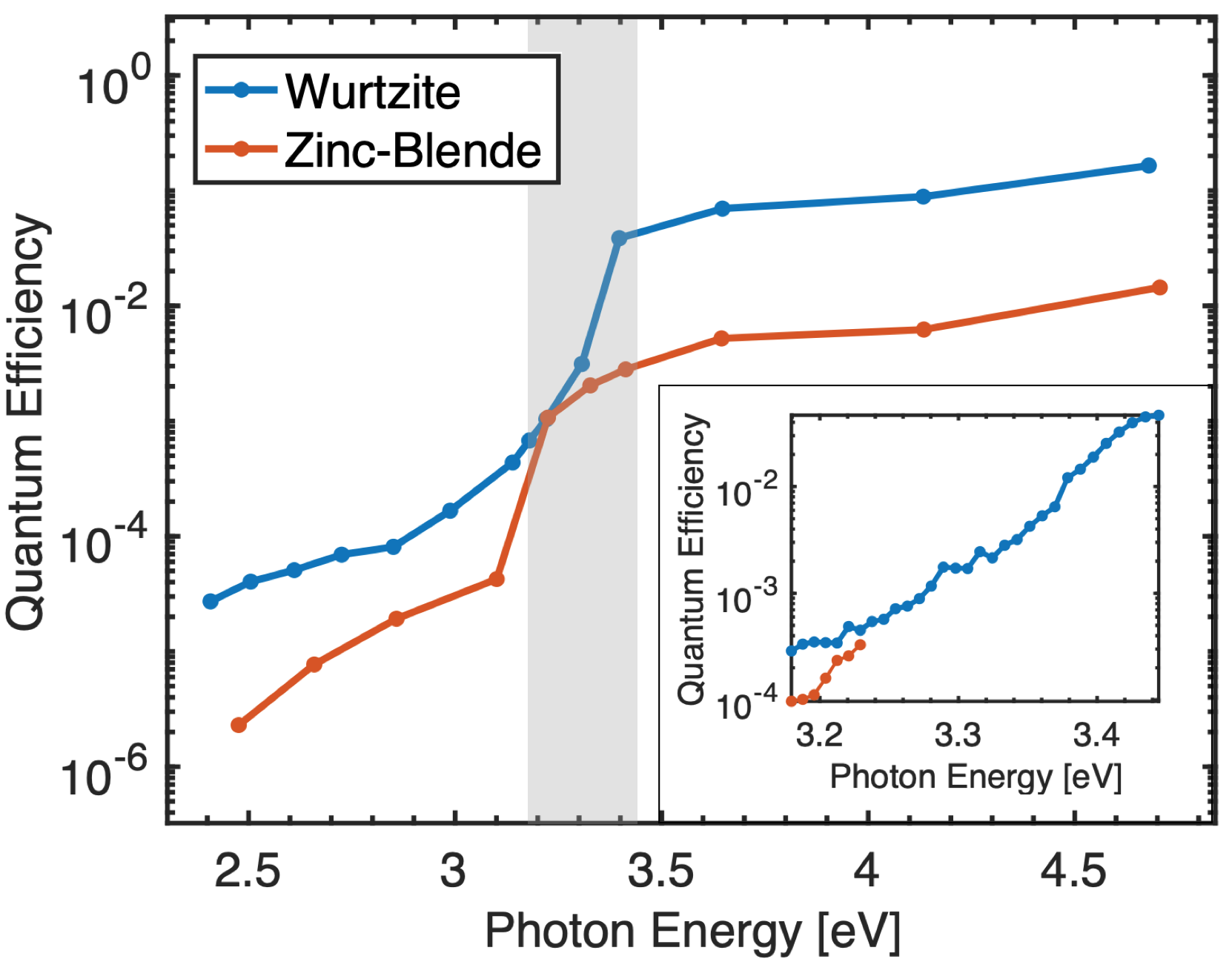}
   \caption{The QE of the ZB and W-GaN photocathodes. The inset data was measured near the spin-polarization region (shown in gray) of the photocathodes with the 1.4 nm bandwidth UV laser. }
   \label{QE}
\end{figure}

Mg-doped (3$\times$10$^{19}$ cm$^{-3}$ Mg concentration) W-GaN and ZB-GaN samples were epitaxially grown on single-crystal W-GaN and 3C-SiC substrates, respectively (see the Supplemental Information for details on sample preparation, narrow band UV laser and spin polarization measurements). They were activated to NEA through Cs deposition yielding QEs of 16.5\% with W-GaN and 5\% with ZB-GaN at 265 nm. To characterize the photoemission, a spectral response was performed with both UV (LEDs) and visible (monochromator with white-light Xenon lamp) light. This data is plotted in Fig. \ref{QE}, where expected bandgap energies of around 3.2 eV and 3.4 eV, and their associated QEs of 0.1\% and 3.9\% (respectively for ZB- and W-GaN) are shown. 

Both the LEDs and monochromator produced light with a typical 10 nm bandwidth (BW). The energy spread corresponding to this bandwidth far exceeds the valence band splitting of W- and ZB-GaN. Consequently, to be able to measure the electron spin polarization (ESP) accurately, a narrow-band (1.4 nm FWHM), tunable UV source was developed.

\begin{figure}[h]
   \centering
   \includegraphics*[width=1.05\columnwidth]{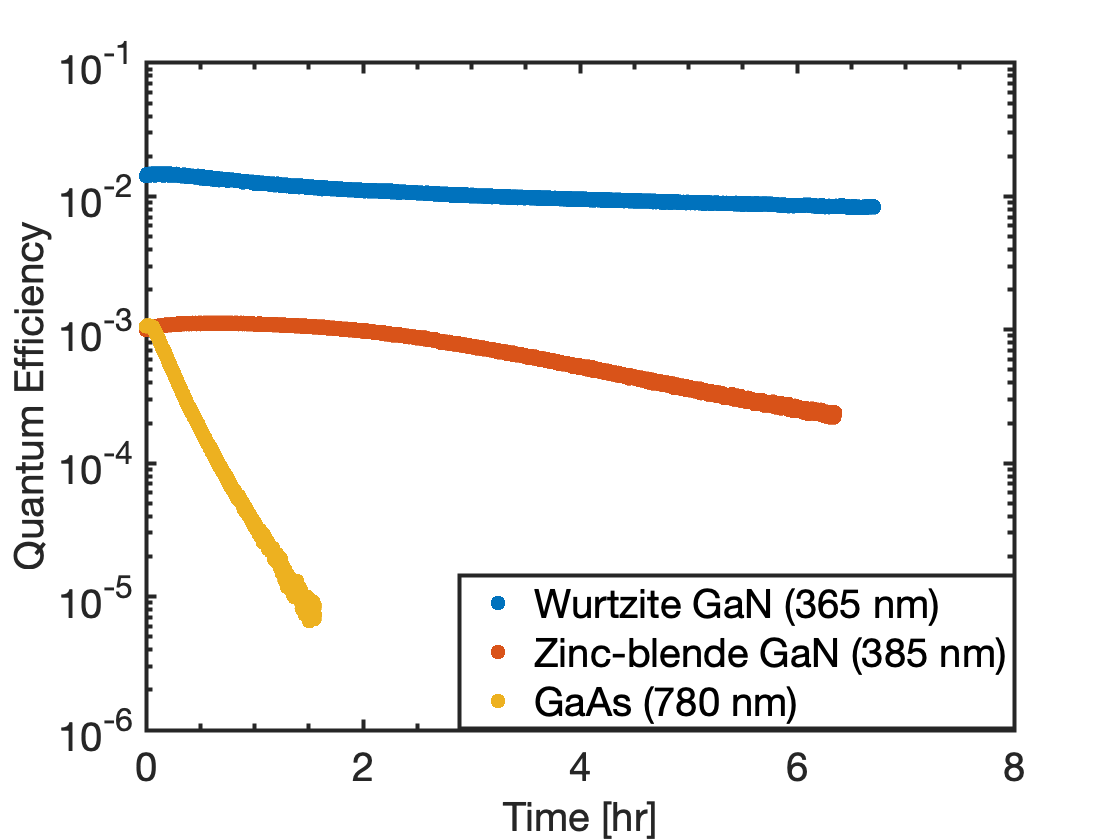}
   \caption{Threshold lifetimes of the NEA W-GaN, ZB-GaN and GaAs photocathodes compared under similar vacuum conditions. }
   \label{lifetime}
\end{figure}

\begin{figure}[h]
   \centering
   \includegraphics*[width=1.06\columnwidth]{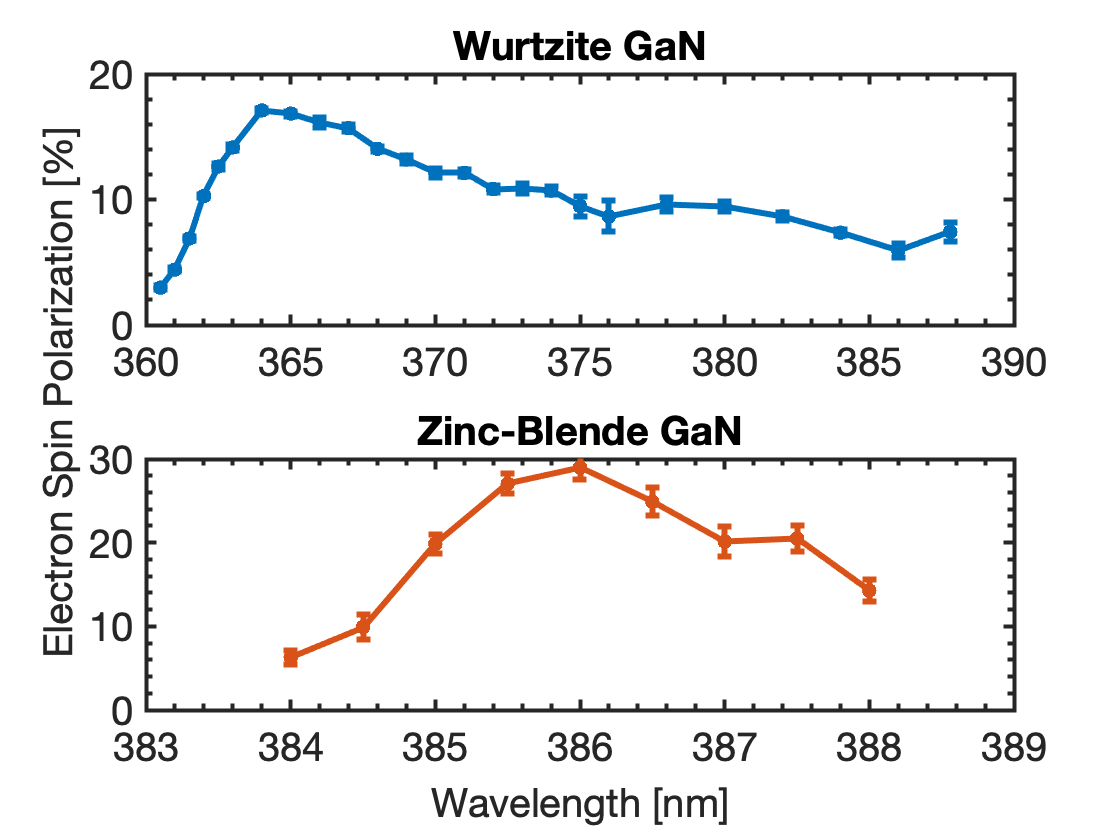}
   \caption{Measured spin polarization from W-GaN (top) and ZB-GaN (bottom) photocathodes. }
   \label{spin}
\end{figure}

To compare dark lifetimes, measurements were performed under similar vacuum conditions at a 2$\times 10^{-10}$ Torr base pressure near each photocathodes' bandgap energy, (365 and 385 nm respectively for W and ZB-GaN) yielding 1/e lifetime values of 11.3 and 4.5 hours, as shown in Fig. \ref{lifetime}. For comparison, a GaAs photocathode (100, Zn-doped at 9 $\times$ 10$^{18}$ cm$^{-3}$ concentration) was activated with Cs only and monitored with a 780 nm laser to measure a corresponding 1/e lifetime. 
As shown in Fig. \ref{lifetime}, we see a lifetime of 0.3 hours, 12 times less than that of the ZB-GaN and 38 times less than that of W-GaN. The higher robustness of NEA GaN photocathodes over GaAs was also found in Ref. \onlinecite{thresholdpaper}. Once spectral and lifetime characterization was complete, the photocathodes were transferred under vacuum into a Mott scattering polarimeter for ESP characterization. The resulting ESP values are shown in Fig \ref{spin}, where ZB-GaN produced a peak ESP of 29\% at 386 nm and W-GaN achieved a peak ESP of 17\% at 364 nm.

\begin{figure}[!h]
   \centering
   \includegraphics*[width=1\columnwidth]{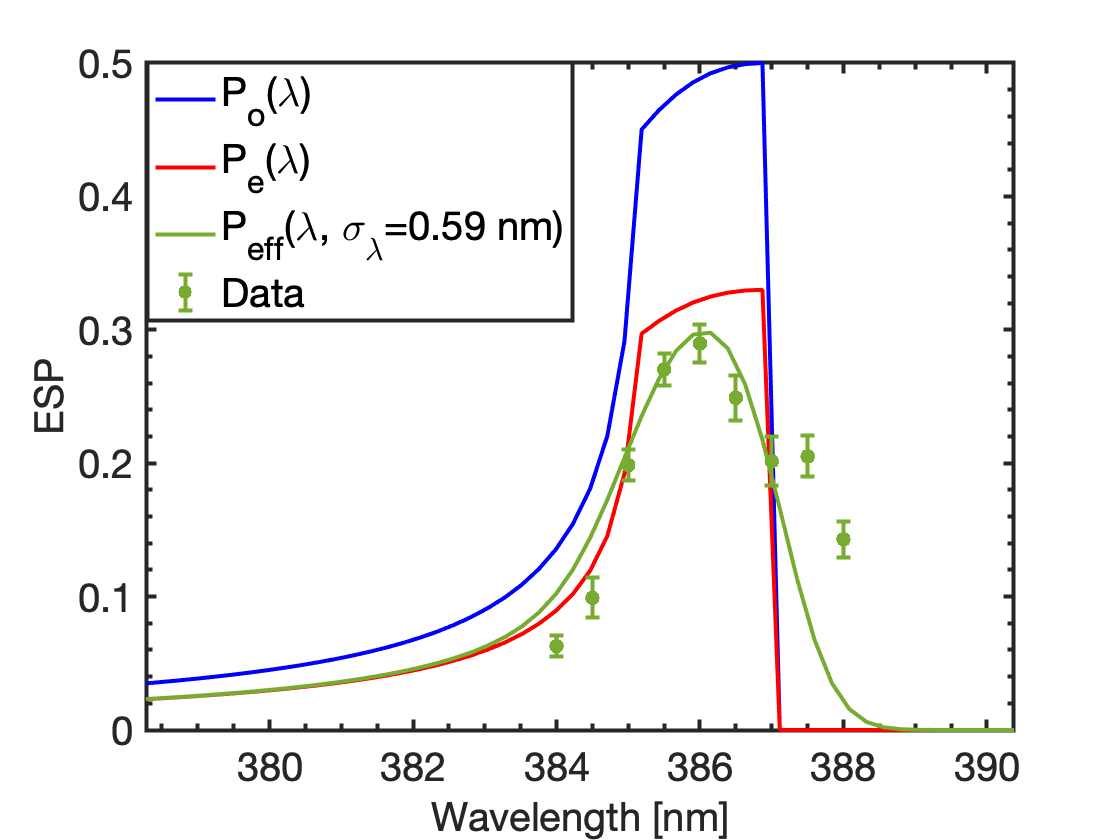}
   \caption{A model to approximate the theoretical spin polarization spectrum of the ZB-GaN photocathode. $P_o(\lambda)$ is the DP \cite{DYP} excitation distribution, described in the text. $P_e(\lambda)$ is the excitation distribution scaled to account for depolarization mechanisms. $P_{\text{eff}}(\lambda, \sigma_{\lambda}=0.59 \text{ nm})$ is the expected curve given 1.4 nm BW incident light.}
   \label{polmodel2}
\end{figure}

To model ZB-GaN spin polarized photoemission, we start with the model from D'Yakonov and Perel \cite{DYP} (DP model), which approximates the spin-polarization of electrons in a ZB structure upon excitement from the circularly-polarized light. This model has also been the foundation of a GaAs Monte Carlo spin polarization calculation \cite{opaper}. ZB-GaN possesses a valence band split between the degenerate HH/LH bands and the split-off (SO) band separated by an energy of $\Delta_{so}$ \cite{valenceboth}. However, in ZB materials with a small $\Delta_{so}$, the SO band states start blending slightly with the HH/LH band states, so this must be considered in spin polarization calculations. Note the DP model cannot be applied to wurtzite structures, which have a non-degenerate valence band. In the model, where $n=1, 2, 3$ correspond to the HH, LH and SO bands respectively, spin polarization is calculated as
\begin{equation}
P_o=\frac{\sum_n P^{(n)}_o K_n}{\sum_n K_n}.
\end{equation}
Here, $P^{(n)}_o$ refers to the initial degree of spin orientation of the electrons excited from each valence sub-band $n$, and $K_n$ are values proportional to the absorption coefficients from the $n$-th valence sub-band. Both $P^{(n)}_o$ and $K_n$ are calculated from the material properties of $E_g$, $\Delta_{so}$, $m_c$, $m_h$ and $m_l$. The last three values are the effective masses of conduction band electrons, heavy-holes and light-holes, respectively. Experimental measurements and calculations typically yield $\Delta_{so}$ in the 15-20 meV range\cite{dsoexp, dsoexp2, dsoexp3, effectivemass} (significantly less than the $\Delta_{so}$ = 340 meV  of GaAs \cite{pierce2}). In terms of $m_0$, we have 
$m_c = 0.15$ \cite{mcpaper}, $m_h = 1.3$ \cite{semiconductorrefbook} and $m_l = 0.19$ \cite{semiconductorrefbook} for ZB-GaN. $P_o(\lambda)$ is shown in blue in Fig. \ref{polmodel2}. Here we see where $P_o(\lambda)$ decreases with decreasing wavelength as the photon energies become large enough to excite electrons from the SO band. The shoulder corresponds to the SO band absorption threshold with an expected 50\% ESP peak.

We can therefore calculate the theoretically expected fraction of spin-up/spin-down electrons computed with the DP model, $N_{\pm}(\lambda)$ from 
\begin{equation}
P(\lambda)=\frac{N_+(\lambda)-N_-(\lambda)}{N_+(\lambda)+N_-(\lambda)},
\label{poldef}
\end{equation}
where $N_+(\lambda)+N_-(\lambda)$ is constrained to be 1. To consider an incident light bandwidth with an associated energy spread comparable to $\Delta_{so}$, we convolve $N_{\pm}(\lambda)$ with light intensity
\begin{equation}
N_{\pm}(\lambda, B) = \int I(\lambda', \lambda, \sigma_{\lambda}) N_{\pm}(\lambda') QE(\lambda') d\lambda',
\label{convolve}
\end{equation}
where $I(\lambda', \lambda, \sigma_{\lambda})$ is the intensity distribution at a wavelength $\lambda$, which in our calculations is assumed to be Gaussian, $\sigma_{\lambda}$ is the bandwidth and $\lambda'$ is an integration variable. In the above formula we also account for the QE dependence on wavelength using data taken from Fig \ref{QE}. Using Eqns. \ref{poldef} and \ref{convolve}, we compute the theoretically expected spin polarization.

The resulting curve ($P_{\text{eff}}(\lambda, \sigma_{\lambda})$) is shown in green in  Fig \ref{polmodel2}, calculated for $\sigma_{\lambda}$ = 0.59 nm (FWHM 1.4 nm). When computing the curve, $E_g$, $\Delta_{so}$ were allowed to vary to best fit the data. We additionally introduce another parameter $M$ as a multiplicative constant to Eqn. \ref{convolve} to account for the electron depolarization that occurs between excitation and emission not accounted in the DP model and assumed to be wavelength-independent. From this model, we obtained fit values of $E_g = 3.205$ eV and $\Delta_{so} = 15$ meV. This $\Delta_{so}$ value agrees with a value reported for a ZB-GaN sample grown on 3C-SiC\cite{dsoexp}. 

The parameter $M$ can be used to scale the DP model curve to approximate the theoretically obtainable ESP with monochromatic light illumination ($P_e(\lambda)$) as shown in red in Fig \ref{polmodel2}. Fitting to our data yielded $M=0.66$, leading to a $33\%$ maximum obtainable ESP. This value is similar to the 35\% typically reported for bulk GaAs, where due to the much larger $\Delta_{so}$, the light bandwidth is ordinarily not considered.

Although ZB-GaN demonstrates higher spin-polarizations, it is much more difficult (with less-developed growth techniques) to grow than W-GaN. Its heteroeptaxial  growth usually produces a defect-rich surface \cite{b3}, specifically full of anti-phase boundary (APB) defects \cite{b5}. APBs and stacking faults originate and propagate up from the interface between the 3C-SiC and GaN due to a substantial lattice mismatch. To characterize these defects, another ZB-GaN sample and another single-crystal W-GaN were grown for atomic force microscopy (AFM) images, shown in Fig. \ref{afm}. Predictably\cite{b5}, the ZB-GaN sample exhibits many APBs where the crystal grain is rotated by 90 degrees. Such defects are not typical of homoepitaxial single-crystal W-GaN samples, reportedly possessing around $10^5$ cm$^{-2}$ dislocation densities \cite{b7, scdislocNEW}, whereas they have been measured around $10^9$ cm$^{-2}$ for ZB-GaN \cite{b5}.

\begin{figure}[!h]
   \centering
   \includegraphics*[width=1\columnwidth]{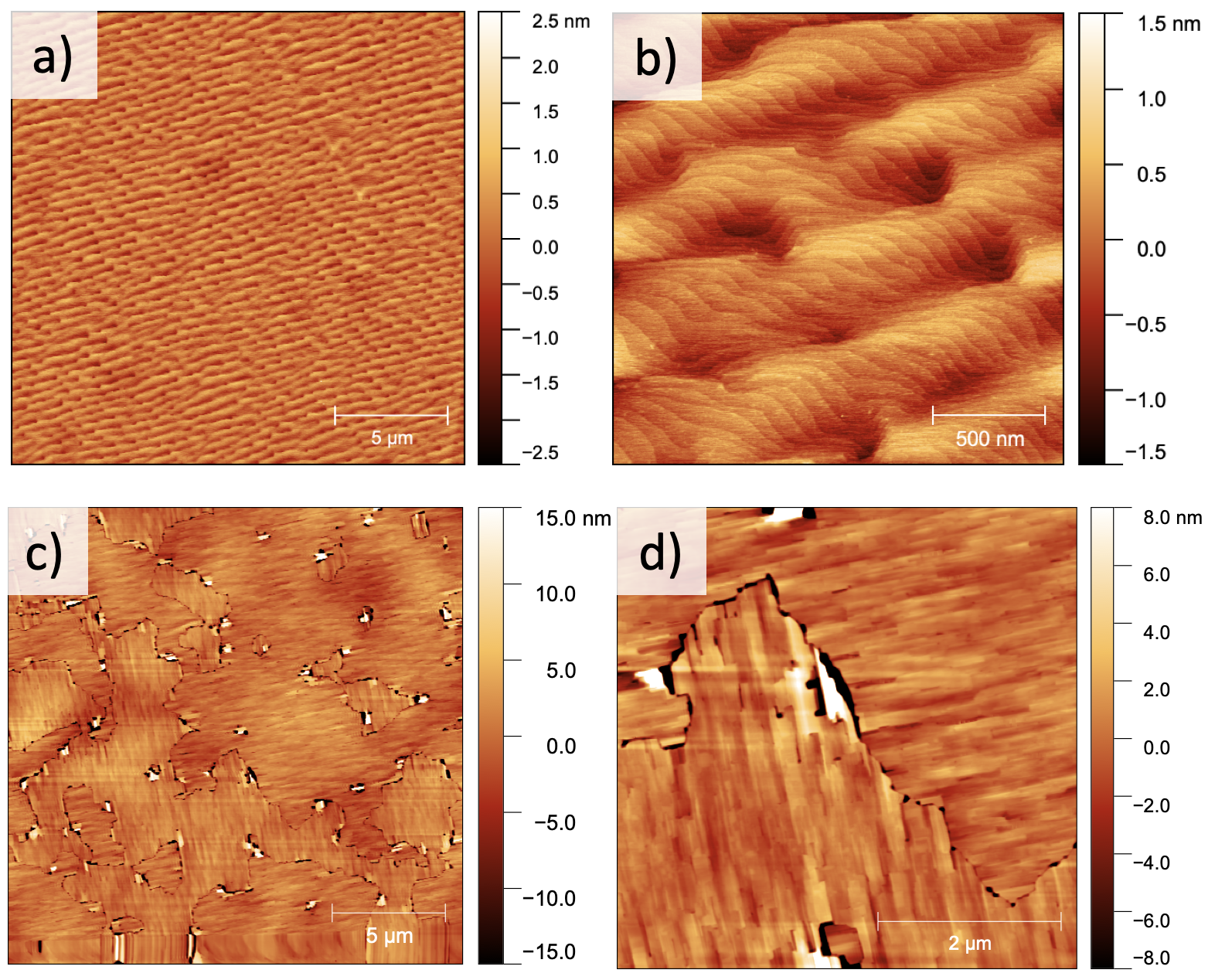}
   \caption{W-GaN and ZB-GaN photocathode AFM scans. The W-GaN photocathode, shown in areas of $a)$ 20$\times$20 $\mu$m$^2$ and $b)$ 2$\times$2 $\mu$m$^2$ has a much smoother surface (0.36 nm RMS roughness) and a consistent crystal grain. The ZB-GaN photocathode, shown in areas of $c)$ 20$\times$20 $\mu$m$^2$ and $d)$ 2$\times$2 $\mu$m$^2$ is characterized by APBs and a relatively rough surface (3.52 nm RMS roughness), as detailed in the text.}
   \label{afm}
\end{figure}


The DP model applied to ZB-GaN can not be applied to a wurtzite structure, where the HH and LH bands are nondegenerate and split from a third valence sub-band called the crystal field (CR) band \cite{kpmethod, valencebandstructure}. The W-GaN HH/LH band separation can be calculated as \cite{kpmethod}
\begin{equation}
    \Delta E = \frac{\Delta_{cr}+\Delta_{so}}{2}-\frac{1}{2}\sqrt{(\Delta_{cr}+\Delta_{so})^2-\frac{8}{3}\Delta_{cr}\Delta_{so}},
    \label{dE}
\end{equation}
where $\Delta_{cr}$ is the crystal field splitting. Given it is non-degenerate, the theoretical maximum W-GaN ESP is 100\%, however, the weak spin-orbit coupling leaves only a very small HH/LH separation. 
The values in Eqn.~\ref{dE} have been extensively calculated and measured \cite{effectivemass,kpmethod, wsosp2,wsosp4, wsosp5, wsosp6, wsosp7, wsosp8, wsosp9, wsosp10, wsosp11}, and a review article on the topic notes experiments seeming to converge around $\Delta_{cr}$ = 10 meV and $\Delta_{so}$ = 17 meV \cite{nitridereview}, yielding a $\Delta E$ = 5.2 meV HH and LH band separation. For these values, a 16.6 meV separation can also be calculated between the LH and CR band\cite{kpmethod}.

W-GaN room temperature spin relaxation time measurements, influenced by substrate characteristics, doping concentration, and dislocation density, span from just under a picosecond to around 100 picoseconds \cite{wrelax1, wrelax2, wrelax3, wrelax4, wrelax5}. The considerably shorter spin relaxation time in W-GaN compared to ZB-GaN\cite{ganspinrev}, which has a room temperature value of around 1 nanosecond\cite{zbrelax1, zbrelax2}, is a possible explanation for lower spin-polarization measured in W-GaN. In one experimental work, a W-GaN sample with a slightly lower dislocation density demonstrated a longer spin relaxation time\cite{wrelax3}. However, a mechanism where an increased dislocation density could lead to an increased spin relaxation time has been investigated \cite{DJ}.

As shown in Fig \ref{QE}, both crystal orientations exhibit photoemission far below the bandgap photon energy, which is common in GaN photocathodes \cite{highqepaper}. However, as shown in Fig. \ref{spin}, the W-GaN photocathode exhibits a long tail in its spin-polarized photoemission with below bandgap photon energies (the top of valence band is at a 3.4 eV photon energy or 365 nm wavelength). We focus our discussion of this unexpected behavior in W-GaN on various defect states found in GaN and on the Franz-Keldysh (FK) effect, both of which have been previously used to explain sub-bandgap photoemission in GaN \cite{energydist,sauty}. 

Nitrogren vacancies and Mg acceptors are common defects found in MBE-grown Mg-doped p-GaN\cite{wganhole1, defectrev}. Nitrogen vacancies have been attributed to bands (seen in various photoluminescence measurements on Mg:GaN) far below (around 0.5 eV and up) the band gap energy \cite{wganphoto1, wganphoto2, wganphoto3, mggapaper, zbganphoto1} and may be responsible for the sub-bandgap photoemission in this region. The primary bands near the bandgap concern Mg acceptor defects \cite{defectrev, mggapaper}. The Mg acceptor ionization energy is slightly over 200 meV in W-GaN, \cite{mggapaper, mggaea1, mggaea2} but only 100 meV in ZB-GaN, \cite{zbganphoto1, zbganmgga}. Mg acceptor states may explain sub-bandgap spin-polarization as they may possess spin-orbit and crystal-field splittings in W-GaN as some works discuss\cite{acceptorpol2, acceptorpol1, acceptorpol3}. In an experimental work, the  P$_{3/2}$ character of the Mg acceptor states was observed via magnetic resonance only in samples with low dislocation density\cite{acceptorpolexp}. Since our W-GaN is such a sample (while the heteroepitaxial ZB-GaN sample has a significantly higher dislocation density), a long sub-valence band spin-polarization tail in the W-GaN sample could be explained by a split Mg acceptor level state. Furthermore, in ZB-GaN, the acceptor level may not be as strongly polarized as in W-GaN because it is not aided by the crystal field splitting present in W-GaN. Mn-doped GaAs may present the same effect, where the Mn acceptor states may also produce spin-polarized photoemission \cite{mngaas1, mngaas2}.

Another possible mechanism is the FK effect, which could also potentially produce spin-polarized electrons from below bandgap photon energies. The FK effect\cite{franz, keldysh}, occurring when strong electric fields (can be externally-applied, though in our case, induced by the dopants) bring some valence band electrons into the bandgap, has also been extensively studied in GaN \cite{ganfk1, ganfk2, ganfk3, ganfk4, ganfk5}. 

Consequently, because of the lower Mg acceptor ionization energy in ZB-GaN discussed above, ZB-GaN typically has higher hole concentrations than W-GaN for the same Mg concentrations. A Hall Effect measurement performed on our W-GaN sample resulted in a $9.1 \times 10^{17}$ cm$^{-3}$ hole concentration, slightly higher than other measurements indicating a $3 \times 10^{19}$ cm$^{-3}$ Mg concentration corresponding with hole concentrations of 1-6 $\times 10^{17}$ cm$^{-3}$ in W-GaN \cite{wganhole1, wganhole2, wganhole3}. Since the 3C-SiC substrate is conductive, parallel conduction prohibited an accurate hole concentration measurement in the ZB-GaN sample, however hole concentrations of $7 \times 10^{18}$ cm$^{-3}$ in ZB-GaN\cite{zbganmgga, zbganhole} are to be expected with our Mg concentration. Considering the ZB-GaN sample likely possesses a higher hole concentration, one would expect it to have a stronger induced electric field, and thus a stronger FK effect than W-GaN. However, as discussed, the dislocation density of the ZB-GaN sample is likely several orders of magnitude larger than that of the W-GaN. Crystal structure defects common to heteroepitaxial ZB-GaN can produce a wide spectra of states around the valence band edge that may not be spin-polarized \cite{b5, buffer}. It is likely the appearance of these defect states drowns out the ability to photoemit spin-polarized electrons in this region, either from a Mg acceptor state or the FK effect.  While Mg acceptor state splittings and the FK effect are interesting mechanisms to explain the wide W-GaN spin-polarization curve, further work is necessary to understand the origins of the broad tail observed in our W-GaN ESP data. 

The demonstration of GaN as a spin-polarized photocathode is a promising result with opportunities for improvements towards the viability for GaN as a robust spin-polarized electron source. ZB-GaN displays spin-polarization comparable to that of bulk GaAs and already boasts a longer dark lifetime. Further lifetime and QE enhancements may come from improved growth techniques to yield a better crystal quality with less defects.  One path towards improved ZB-GaN films is through the use of an AlN nucleation layer by improving the lattice mismatch between GaN and 3C-SiC, thereby reducing crystal defects \cite{buffer,aln}. 

A possible path towards improving the ESP of ZB-GaN is to introduce crystal strain to remove the HH/LH degenerancy, commonly done with GaAs. The properties of strained W-GaN are already well-understood \cite{kpmethod, wsosp11, straintheory2, straintheory3}, and works detail experiments done with both strained W-GaN \cite{strainexp1} and ZB-GaN structures\cite{slgan}.  One work reported a nearly 100\% circularly polarized photoluminescent beam produced with such a sample \cite{strainedpol1}.

While the ability to grow W-GaN with superb crystal quality already yields high QEs and long dark lifetimes, it has a relatively low ESP value. Cooling W-GaN during photoemission may improve the ESP value, as one work shows that the spin-relaxation time in n-type GaN increases by a factor of 5 at 50 K from that of room temperature\cite{relaxtemp}. Lowering the doping concentration may also help increase the spin relaxation time, although doing so could sacrifice QE.

W- and ZB-GaN QE may be improved with the use of Distributed Bragg Reflectors (DBRs), as done with GaAs\cite{gaasdbr1, gaasdbr2}.  DBRs have been grown with both W-GaN \cite{dbrgan1, dbrgan2, hgxdbr} and ZB-GaN\cite{cubicdbr1}. 
Further photocathode robustness improvements may also be found through robust hetero-structures used as NEA activation layers, which have been developed to improve the operational lifetime of GaAs\cite{j1paper, j2paper, j3paper, j4paper, j5paper}, and through enhancements in N-polar GaN\cite{npolar2}, which has demonstrated a Cs-free NEA condition\cite{Npolar1}. The GaN bandgap may be tuned with indium incorporation (InGaN)\cite{ingan1, ingan2, ingan3, inganpol}, which may provide a path towards an incident-light-tunable spin-polarization source. 

Spin-polarized photoemission has been measured for the first time from both wurtzite and zinc-blende GaN photocathodes. The p-doped photocathodes were epitaxially grown and characterized in a Mott Polarimeter. The wurtzite GaN photocathode exhibited a higher quantum efficiency and dark lifetime at threshold wavelengths than that of the zinc-blende photocathode. However, using a tunable OPA, a peak spin polarization of nearly twice that of wurtzite GaN was observed in zinc-blende GaN.  

See the Supplementary Information for additional details on GaN sample growth and the narrow-band circularly polarized UV light source.

The authors would like to thank J.K. Bae, N. Otto and M.A. Reamon for experimental assistance and T. Arias, L. Cultrera, O. Chubenko and T. Wu for thoughtful discussion.  This work is supported by United States Department of Energy (DOE) grant DE-SC0021002 and by the U.S. National Science Foundation under Award PHY-1549132, the Center for Bright Beams.

\section*{Data Availability Statement}
The data that support the findings of
this study are available from the
corresponding author upon reasonable
request.

\section*{Supplemental Information}

\section{Sample Preparation}

GaN samples were grown in a Veeco Gen10 molecular beam epitaxy (MBE) system at Cornell University. The chamber features effusion cells filled with elemental sources of gallium, aluminum and magnesium. To grow W-GaN, a 2$\times$2 in$^2$ single-crystal GaN substrate was purchased from Ammono and diced into a 7$\times$7 mm$^2$ square. For the ZB-GaN substrate, a 4$\times$4 in$^2$ wafer of a 10.3 $\mu$m thick silicon carbide (3C-SiC) layer on a 525 $\mu$m thick silicon (Si 100) base purchased from NOVASiC was diced into a 1$\times$1 cm$^2$ square. The surface morphologies of the two substrates are shown in atomic force microscopy (AFM) scans in Fig. 1. Prior to sample growth, the substrates were ultrasonicated sequentially for 10 minutes each in acetone, methanol and isoproponol before being inserted into the MBE chamber, where they were outgassed at 200 degrees Celsius for 7 hours. Next, a 100 nm thick unintentionally-doped (UID) GaN layer was grown on each to set the epitaxial, atomic (W or ZB) structure of the material. This was then followed by a 1 $\mu$m thick Mg-doped GaN layer, which was grown in each respective orientation. To optimize the QE of the photocathodes, the doping concentration was set to 3$\times10^{19}$ cm $^{-3}$ based on work reported elsewhere \cite{highqepaper}. Metal-rich conditions were maintained throughout the epitaxial growth process, grown within the step-flow growth mode. Ga droplets were desorbed in-situ at the conclusion of the growth.

\begin{figure}[!h]
   \centering
   \includegraphics*[width=1.05\columnwidth]{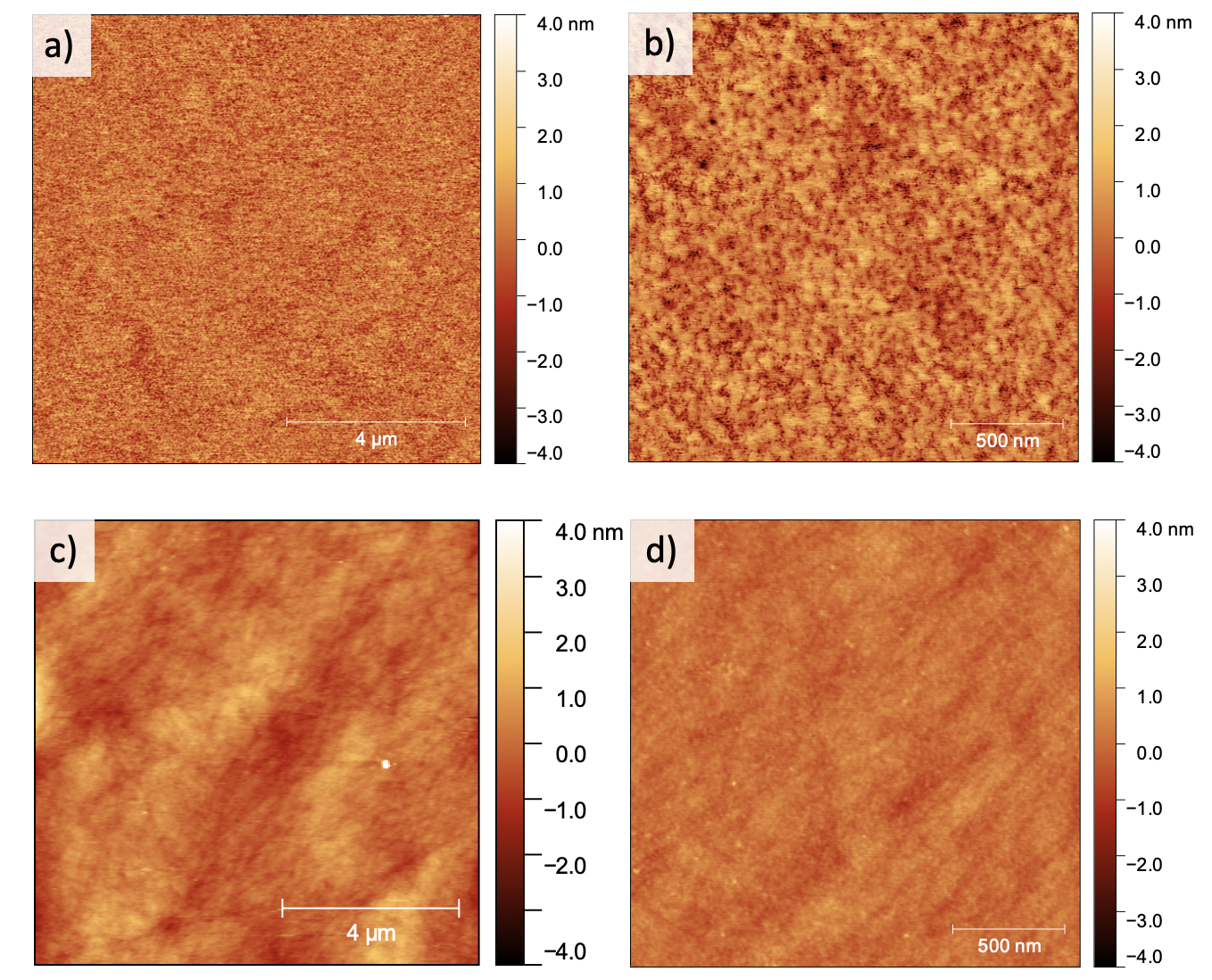}
   \caption{AFM scans of the single-crystal W-GaN substrate and 3C-SiC substrate. The single-crystal W-GaN substrate, shown in areas of $a)$ 10$\times$10 $\mu$m$^2$ and $b)$ 2$\times$2 $\mu$m$^2$ has an RMS roughness of 0.57 nm, while the 3C-SiC substrate, shown in areas of $c)$ 10$\times$10 $\mu$m$^2$ and $d)$ 2$\times$2 $\mu$m$^2$ has an RMS roughness of 0.49 nm.}
   \label{afm}
\end{figure}

After growth, the samples were removed from the MBE system and placed into a static vacuum suitcase (exposed to air for less than 10 seconds) to be transported across campus to the Bright Beams Photocathode Laboratory \cite{bblpaper}. Once transported, the suitcase was vented and the sample was mounted on a stainless steel puck. The sample was fixed to the puck by a cap with a 7 or 5 mm diameter hole to expose the GaN. For good electrical and thermal contact, a small amount of indium foil was placed on each of the 4 edges of the sample beneath the cap. During this process, the photocathode was exposed to air for approximately five minutes before insertion into the photocathode activation chamber.

The samples were annealed at 600 degrees Celsius for 6 hours prior to NEA activation. Here, the photocurrent was monitored with 10 $\mu$W of 265 nm light generated by an LED incident on the photocathode while Cs was deposited onto the surface with a filament source. Cs deposition was stopped when the QE peaked.

 \section{Narrow band UV Source}
Characterization of a GaN photocathode as a spin-polarized electron source requires illumination with a narrow spectral bandwidth due to the relatively small valence band splitting.  Additionally, for spectral analysis of both W and ZB-GaN, broad tunability between 360-390 nm is required. To accomplish this, we constructed an optical parametric amplifier (OPA) with output in the near-infrared and added a frequency-doubling stage to achieve the required UV wavelengths.  The OPA is powered by a 20W PHAROS laser system purchased from Light Conversion which generates approximately 100 fs laser pulses at a repetition rate of 1.1 MHz and a wavelength of 1030 nm. The beam is split between two paths. In the first path, the light is frequency doubled to 515 nm and used as the pump source of the OPA. In the second path, the pulses are tightly focusing into an undoped YAG crystal to generate a (chirped) white light supercontinuum for the OPA signal beam. The 515 nm and white light are then recombined inside a beta barium borate (BBO) nonlinear crystal where the OPA process occurs. A tunability range from 720–780 nm was achieved by adjusting the crystal orientation and relative arrival times of the signal and pump light inside the BBO crystal. 

The OPA output is frequency-doubled using a bismuth triborate (BiBO) nonlinear crystal \cite{bibo}. Over the OPA tunability range, the frequency doubled process typically had 5-10\% conversion efficiency, nominally producing 10-20 mW of narrow-band UV light. The beam is then launched into a single-mode UV optical fiber for transport to the Mott polarimeter. Nonlinear spectral broadening observed in the fiber was minimized by reducing the laser intensity prior to launching into it. At the sample, we achieved a spectral bandwidth (full-width half-maximum) of 1.4 nm and a power level of 5 $\mu$W across the tunable UV ranges.

Circularly polarized light is required to excite an electron beam with a net spin polarization. Upon exiting the fiber, the light is elliptically polarized at an arbitrary angle. Circular polarization was obtained using a linear polarizer and a quarter-waveplate with the fast axis oriented 45 degrees off the polarizer. To account for bias in the detection rates, count rates are measured for both left- and right-handed light during the measurement. Handedness was switched by rotating the waveplate 90 degrees.

 Prior to measurements with the Mott polarimeter, the degree of circular polarization was measured. A second linear polarizer was placed after the waveplate, and transmitted optical power was monitored while rotating the second polarizer through a full 360 degrees. The degree of circular polarization was calculated using\cite{piercepaper}
\begin{equation}
    P_{cp} = \frac{2(I_{min}I_{max})^{1/2}}{I_{min}+I_{max}},
\end{equation}
where $I_{min}$ and $I_{max}$ are the minimum and maximum intensities achieved during rotation. The optimal waveplate angles for the $\lambda/4$ and $3\lambda/4$ were found such that the degree of circular polarization for each handedness was 99.8\%.

\section{Spin Polarization Measurements}
The spin polarization measurements were performed in a Mott Scattering Polarimeter with a 20-keV retarding bias on a tungsten target\cite{mottpaper, jlabmottpaper}. A 90 degree electrostatic bend rotates the initially longitudinally oriented electron spin to the transverse plane as defined by the beam direction. Additional electrostatic correctors are used to guide the beam to the target and Einzel lenses provide beam focusing. Scattered electrons are registered in two symmetrically placed Channeltron detectors placed on both sides of the target. A Sherman function value of 0.18 was calibrated in an earlier work \cite{j3paper}.
 
 After circularly polarized light was produced as detailed above, it was aligned to the photocathode and the electron optics for beam transport to the target were tuned to optimize the count rates on the Channeltron detectors. The ESP value was determined by measuring the count rate asymmetry, $A$, between detectors \cite{jlabmottpaper, dunning}, 
\begin{equation}
     A = \frac{\sqrt{L_-R_+}-\sqrt{L_+R_-}}{\sqrt{L_-R_+}+\sqrt{L_+R_-}},
\end{equation}
where $L$ and $R$ correspond to the integrated number of counts registered on the left and right detectors, respectively, while $+$ and $-$ denote the two handedness states of the light. The measured spin-polarization is the asymmetry scaled by the effective Sherman function. At each wavelength, 20 measurements were taken and averaged. Background detection rates were registered and subtracted at each step of the measurement. The reported uncertainties are statistical and do not account systematic errors in the instrument.

\bibliography{sources.bib}

\end{document}